# Nonreciprocal optical circuit switching

Zhifeng Tu[1,2,†], Yucong Yang[3,4,5,†], Yiran Wei[2,†], Shuyuan Liu[2,*], Fangchen Hu[2], Peng Zou[2], Chengkun Yang[2], Tianchi Zhang[3,4,5], Di Wu[3,4,5], Ruoyu Shen[2], Bingzhou Hong[2], Haiwen Cai[2], Lei Bi[3,4,5,*], and Wei Chu[2,*]

[1]College of Future Information Technology, Fudan University, Shanghai 200433, China

[2]Zhangjiang Laboratory, Shanghai 201210, China

[3]National Engineering Research Center of Electromagnetic Radiation Control Materials, University of Electronic Science and Technology of China, Chengdu 610054, China

[4]State Key Laboratory of Electronic Thin-Films and Integrated Devices, University of Electronic Science and Technology of China, Chengdu 610054, China

[5]Key Laboratory of Multi-Spectral Absorbing Materials and Structures of Ministry of Education, University of Electronic Science and Technology of China, Chengdu 611731, China

†These authors contributed equally to this work

*[liusy@zjlab.ac.cn](liusy@zjlab.ac.cn) (S.L.), [bilei@uestc.edu.cn](bilei@uestc.edu.cn) (L.B.), [chuwei@zjlab.ac.cn](chuwei@zjlab.ac.cn) (W.C.)

**Abstract:** Directly switching optical signals outperforms conventional optoelectronic hardware in terms of cost, latency, and energy efficiency, and is expected to address the growing demand for data node capacity driven by the development of machine learning and artificial intelligence (AI) technologies. Therefore, optical circuit switching (OCS) technology has piqued widespread research interest in various technical solutions, including silicon photonics. However, silicon-based integrated OCS remains constrained by challenges such as network performance and port scalability. Here we propose a magneto-optical heterogeneous integrated nonreciprocal OCS (NOCS) network based on a silicon photonics platform, achieving bidirectional full-duplex nonreciprocal transmission by programming reciprocal and nonreciprocal phase shifters. We demonstrate that compared with the existing OCS architecture, NOCS has the advantages of ultra-high reconfiguration speed, large-scale integration compatibility, and bidirectional channel isolation reducing the number of required ports. NOCS could meet the programming speed requirements of the AI backend network, or supports nonreciprocal optical switching applications without multiplexing technology.

## 1. Introduction

In recent years, with the explosive development of machine learning and artificial intelligence (AI) technologies[1, 2], the node traffic demand in AI data centers has reached an unprecedented scale, particularly in AI backend networks such as memory-to-GPU and GPU-to-GPU interconnects. Data transmission in data centers relies heavily on high-speed optical fibers between servers, racks, and network switches, where optical signals are received by optical transceivers within the switches and converted into electrical signals for electronic switching[3]. While this optical-electrical-optical (O-E-O) conversion-based approach has enabled data centers to achieve high throughput and flexibility, it faces significant bottlenecks in power consumption and latency as traffic demands continue to grow exponentially with the rise of AI and cloud computing[4, 5].

Optical circuit switching (OCS) provides low power consumption[6, 7], low latency, and seamless compatibility with diverse bandwidth requirements, serving as a pivotal technology for next-generation optical networks[7, 8]. OCS eliminates the need for power-intensive O-E-O conversions, significantly reducing energy consumption. OCS also offers ultra-low latency since optical signals bypass the electronic processing bottleneck, enabling near-instantaneous reconfiguration speed[8, 9, 10]. Furthermore, OCS is inherently compatible with high-bandwidth optical fibers, making it an ideal candidate for scaling data center networks to meet future demands. Google utilizes the Apollo OCS system in their data center, achieving a 30% increase in throughput while reducing power consumption by 40%, demonstrating the practical advantages of OCS in large-scale AI workloads[11, 12].

Among various OCS technologies, such as micro-electromechanical systems (MEMS)[13, 14], liquid crystal optical switches[15, 16], and piezoelectric beam control[17]; silicon photonic switching technology stands out in high-speed switching scenarios due to its ultra-fast reconfiguration speed, which can reach the nanosecond scale.[7, 18]. Furthermore, silicon photonics enables the integration of other optical components, such as light sources and optical amplifiers on OCS chips[7, 19, 20, 21, 22], making it possible to realize completely lossless and fully functional optical switching chips.

To realize silicon-based integrated optical switching devices, several technical approaches have been explored, including electro-optic effect[23, 24, 25], thermo-optic (TO) effect[26, 27, 28], free carrier dispersion[29, 30], and silicon-integrated MEMS[31, 32, 33]. Recently, optical switches based on magneto-optical (MO) effects have attracted extensive research attention owing to their process compatibility with silicon-based heterogeneous integration, nanosecond switching speed, and unique nonreciprocal transmission characteristics[34, 35, 36, 37]. Despite their potential, MO OCS networks have not yet been reported. Nevertheless, the ultra-high switching speed and nonreciprocal characteristics of these switches offer significant potential in small-scale all-optical switching networks[36]. These features not only meet the reconfiguration speed requirements but also solve the isolation problem during bidirectional transmission of the same wavelength signal in a single optical path, thereby reducing the number of required ports or achieving full-duplex operation for uplink and downlink[38, 39, 40, 41, 42].

This work demonstrates a nonreciprocal OCS (NOCS) network that is potentially capable of large-scale integration and supports bidirectional independent transmission for the first time. We realized a silicon-based integrated MO NOCS network on silicon-on-insulator (SOI) wafers by cascading MO Mach-Zehnder Interferometer (MZI) switches to form a 4 × 4 Benes network using monolithically deposited Ce:YIG films on silicon. The device architecture is shown in Fig. 1a, and the structure of the MZI switch unit is shown in Fig. 1b. Besides the traditional reciprocal phase shifter (PS), the device is also equipped with push-pull nonreciprocal PSs, allowing the forward and backward switching states of the unit devices to be completely decoupled, i.e., bidirectional transmission independently exhibits "bar" or "cross" states. Specifically, it supports four operating states: Cross-forward and Cross-backward (CC), Bar-forward and Bar-backward (BB), Bar-forward and Cross-backward (BC), and Cross-forward and Bar-backward (CB), thereby fulfilling the transmission logic requirements of bidirectional independent full-duplex networks. Therefore, in addition to realizing the traditional switching function, the proposed OCS network can also be configured into multiple nonreciprocal operation functions. Figure 1c indicates the network nonreciprocal transmission logic when the middle (two sides) column unit devices are in CB (CC) state, and demonstrates that the network exhibits a unique nonreciprocal transmission matrix.

To fabricate the designed network, we developed an integration process to precisely control the on-chip distribution of MO materials, which can effectively reduce the unnecessary absorption loss and limit the contamination of metal elements to other on-chip devices (see Methods section for details). The fabricated cross-section of the MO nonreciprocal PS is shown in Fig. 1d, and the distribution of $H_x$ field components of the $TM_0$ mode propagating in it is shown in Fig. 1e. NOCS can be applied in various scenarios, such as scale-up and scale-out networks in data centers, and transceiver links for integrated sensing and communication (ISAC), to realize multi-port circulation, uplink & downlink full-duplex and channel isolation as shown in Fig. 1f. After explaining the operation principles of the proposed networks, the application advantages in corresponding scenarios will be discussed in detail.

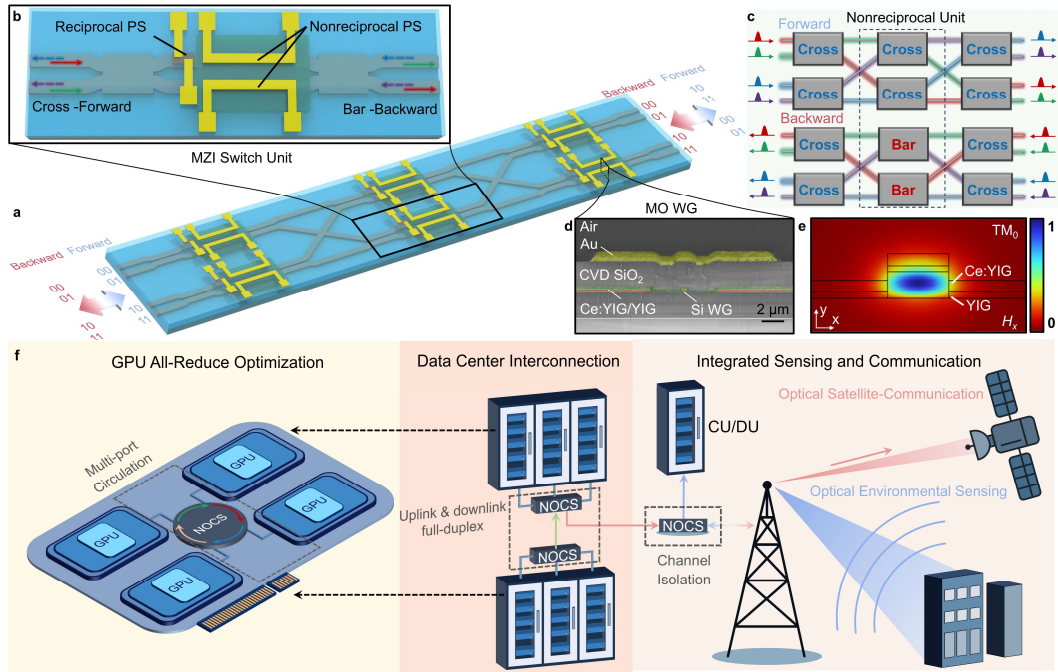

**Fig. 1 The proposed MO NOCS. a**, Schematic of the Benes network architecture consisting of unit MZI optical switches. **b**, Schematic of the four-state MO switch unit. **c**, One nonreciprocal state of the NOCS when the middle (two sides) column units are in CB (CC) state. **d**, Cross-sectional view of the designed MO PS. **e**, Distribution of $H_x$ field components of $TM_0$ mode in MO waveguide (WG). **f**, application of NOCS: All-reduce optimization in on-chip systems; uplink & downlink full-duplex transmission in data centers; and channel isolation in ISAC systems.

## 2. Magneto-optical switch performance characterization

We started with a single 2 × 2 MO nonreciprocal switch on a silicon platform. Figure 2a shows the proposed MO switch, comprising an MZI with two 3-dB multimode-interference (MMI) couplers at both ends. Each arm of the MZI contained an MO/Si WG that functioned as a nonreciprocal PS for the transverse magnetic (TM) mode (MO material characterization can be found in Supplementary Sections 1&2). In a push-pull configuration, the two gold electrodes on both arms were supplied with equal and opposite currents, generating magnetic fields that magnetize the MO films to induce a nonreciprocal phase shift (NRPS). Figures 2b and c illustrate the magnetic field and thermal field distributions around the electromagnet simulated by COMSOL Multiphysics, respectively. The simulations confirmed that the electrodes could provide a sufficient magnetic field to magnetize the MO material to the desired state, but also introduced a significant heat generation, which could lead to degradation of the Faraday rotation (FR) coefficient of the MO materials, as shown in Figure 2d. The MO performance of the film at 80 °C was ~73% of that at room temperature. By trading off the magnetic and thermal effects, a 1500-μm-long MO PS could generate a π/2 NRPS under 400 mA current inputs with a magnetic field of ~90 Oe and a temperature of ~340 K above the WG. Additionally, the lower arm of the MZI incorporated a TO PS with a Ti micro-heater to provide a designed thermal field distribution (as shown in Fig. 2e inset) and a reciprocal phase shift (RPS) to calibrate the operation wavelength. Figure 2e illustrates that the reciprocal PS could achieve a sufficient phase shift of π under 2.75 V input. Furthermore, the RPS exhibited a quadratic dependence on the applied voltage, in excellent agreement with both theoretical and measured results.

To investigate the phase shift capability of the MO PS, the forward and backward transmission spectra between Ports$_{1\&3}$ are shown in Fig. 2f, where a current in the integrated electromagnet was swept from positive (+$z$ in upper arm, −$z$ in lower arm) to negative values and subsequently reversed back to positive. As the current swept from positive to negative values, the interference peaks in the forward (backward) transmission spectra underwent a redshift (blueshift). Conversely, when the current swept back from negative to positive values, the interference peaks in the forward (backward) transmission spectra exhibited a blueshift (redshift), returning to the original positions. Such highly symmetrical shifts in the spectra proved that the TO phase shifts caused by the electromagnets in the two arms basically canceled each other out, making the interference peak shift almost dominated by the MO effect. Further details can be found in Supplementary Section 3. Figure 2g presents the NRPS corresponding to the current sweep in Fig. 2f, exhibiting hysteretic behavior. As the current approached −400 mA (+400 mA), the NRPS continued to decrease (increase), indicating that the MO film remained unsaturated to prevent thermal decay.

We characterized the dynamic response of the MO switch using a measurement setup in which oscilloscope acquisition was synchronized with the output signal of an arbitrary waveform generator (AWG). Figure 2h shows the dynamic response of the designed switch when the electromagnets were driven by a 2 MHz square-wave current pulse. The measured 10%–90% rise and fall times were 58 ns and 57 ns, respectively, demonstrating the high-speed switching capability of the proposed device. Further experimental details are provided in Supplementary Section 4.

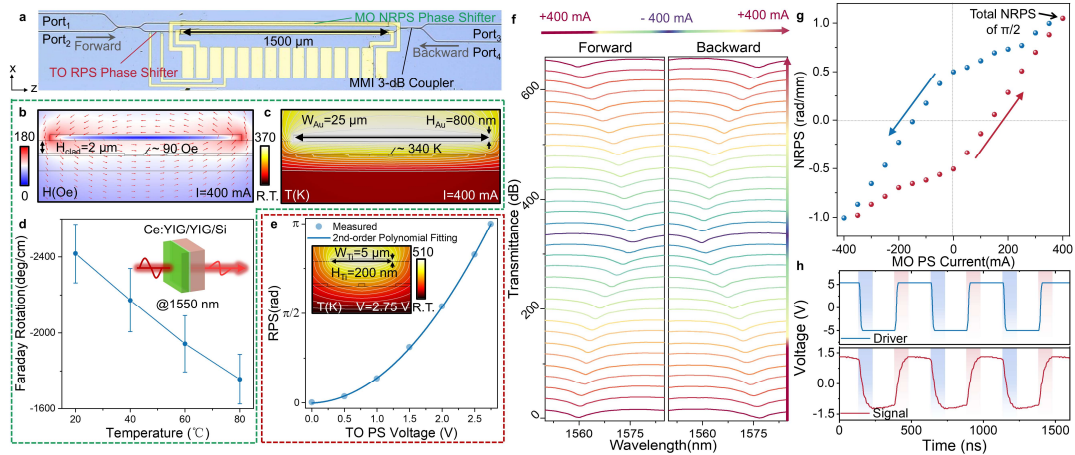

**Fig. 2 Characterization of the MO switch. a**, Optical microscope image of the fabricated MO switch with 1500-μm-long MO PSs. (**b**) Magnetic field and (**c**) thermal field distribution in the nonreciprocal PS cross-section when a current of 400 mA is injected into the electromagnet. **d**, FR coefficient degradation curve with increasing temperature based on free space testing. **e**, RPS as a function of the applied TO PS voltage. The inset shows the thermal field distribution in the reciprocal PS cross-section when a voltage of 2.75 V is applied to the Ti micro-heater. **f**, Forward (left) and backward (right) transmission spectra, with the current swept sequentially from positive to negative and then reversed. **g**, Hysteresis of the device NRPS calculated from **f**. **h**, Time-domain response of the MO switch.

## 3. Transmission logic of a switch unit

To enable the expected transmission logic in the nonreciprocal network, we next demonstrated the switching functionality of the single MO unit device by leveraging its reciprocal and nonreciprocal PSs. Since network testing has higher requirements for device loss and power consumption, we shortened the MO PS to 950 μm and designed it to generate a $\pi/4$ NRPS under 350-mA input, thereby reducing insertion loss (IL) and Joule heat generation by sacrificing crosstalk performance. Figure 3a depicts the four-state MO switch serving as a unit device of the MO network. The RPS produced by the TO PS and the NRPS imparted by the MO PS jointly realized the four transmission states of the MO switch, with specific details in Supplementary Section 5.

In Figs. 3b and c, the magnetic fields generated by currents of equal magnitude and same direction result in identical magnetization orientations of the MO films in the two arms, inducing the NRPS to be 0, causing the unit switch to operate as a reciprocal device. The state in Fig. 3b was defined as the CC with an RPS of 0. In Fig. 3c, the RPS was adjusted to $\pi$, thereby realizing the BB state. In Figs. 3d and e, applying opposite currents to the two-arm nonreciprocal PSs brought a designed NRPS of $\pm\pi/4$. The RPS was appropriately adjusted to $\pi/2$ to compensate for phase differences in forward and backward transmissions, achieving the BC and CB states.

The transmission matrices of the unit device are shown in Figs. 3f-i at the wavelength 1564.84 nm. In the reciprocal operating states, the IL and crosstalk of the device were 7.5–7.9 dB and < −17 dB. In the nonreciprocal operating states, the IL of the device was 8–8.7 dB, and the crosstalk ranged from > −9.6 dB. Notably, the device exhibited significantly better crosstalk performance in the reciprocal states compared to the nonreciprocal states due to the insufficient NRPS, which can be remedied in the future by employing magnets to generate the magnetic field on chip, eliminating FR degradation caused by the thermal effect from the electromagnet[43, 44, 45, 46], achieving an ideal crosstalk of < −20 dB under all four operating states.

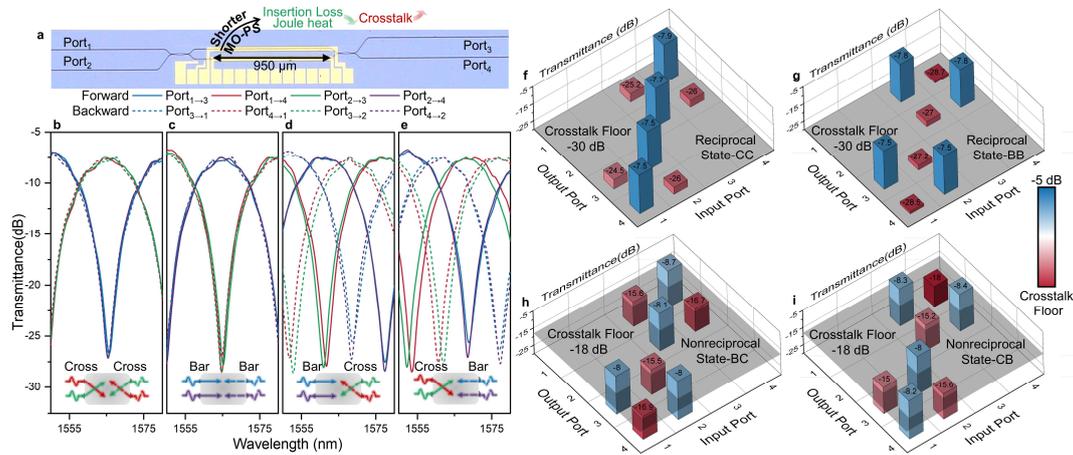

**Fig. 3 Experimental results of the four-state MO switch unit. a**, Optical microscope image of the fabricated four-state switch with a 50-μm-long TO PS and 950-μm-long MO PSs. **b**–**e**, Spectral response of the four-state switch in the state of CC (**b**), BB (**c**), BC (**d**), and CB (**e**). **f**–**i**, Measured transmission matrices of the four-state switch in the state of CC (**f**), BB (**g**), BC (**h**), and CB (**i**) at the wavelength of 1564.84 nm.

## 4. Nonreciprocal reconfigurable OCS networks

Finally, we cascaded the proposed four-state MO switch units to establish a three-layer 4 × 4 nonreciprocal network based on Benes topology architecture. Figure 4a shows the microscopic image of the fabricated MO network, comprising six switch units. Monitoring the split light from the calibration ports, we employed the reciprocal PSs to align the interference peak of the switch units at the operation wavelength (see Supplementary Sections 6&7 for more details). Equipped with both reciprocal and nonreciprocal PSs for the implementation of four-state switching, the 8-port Benes switch can be configured into 4096 ($4^6$) global switching states corresponding to 576 (4! × 4!) permutations; the reconfiguration flexibility was higher than the traditional OCS in the same scale with 64 ($2^6$) global switching states, corresponding to 24 (4!) permutations. This architecture supports various routing strategies to dynamically reconfigure the 4 × 4 MO network into either reciprocal or nonreciprocal states with diverse functionalities.

The routing configurations of the 4 × 4 MO switch are direction-dependent and can be independently manipulated in the forward and backward directions. Herein, we analyze one of the configurations of NOCS as shown in Figs. 4b and c. In the forward direction, the input signals from Ports L1–4 were routed to Ports R1, R4, R3, and R2, respectively; in the backward direction, the input signals from Ports R1–4 were routed to Ports L3, L4, L1, and L2, respectively. The IL of this nonreciprocal OCS ranged from 21 dB to 23.7 dB, with crosstalk > −9.6 dB at a wavelength of 1576.70 nm. It is worth mentioning that the device performance has not yet reached its optimum, which is primarily attributable to the insufficient NRPS of the switch unit. Additionally, fabrication imperfections and non-uniformities in the wafer or etching process should be taken into consideration as contributing negative factors. Nonetheless, the results demonstrate the promising potential of the MO nonreciprocal network in OCS applications.

The communication capability of the 2 × 2 NOCS was experimentally demonstrated, as shown in Fig. 4d corresponding to GPU switching and all-reduce optimization scenario. The capacity demands of PCIe and PON were verified and found to be potentially suitable for data centers and access networks. The signal-to-noise ratios (SNRs) of the reciprocal and nonreciprocal paths at a 64 Gbps data rate were 9.49 dB and 8.39 dB, respectively. The measured eye-diagrams are shown in the insets. To show the scalability of the proposed architecture, the most ideal performances of the NOCS at 2 × 2, 4 × 4 and 8 × 8 port-scales, based on the proposed switch unit were simulated with parameters of real components and devices, as shown in Figs. 4e and f, where the data rate was defined as the maximum error-free capacity without any digital signal processing. The receiver sensitivity of the 8 × 8 NOCS deteriorated severely due to the higher amplified spontaneous emission (ASE) noise of the EDFA, since higher gain was needed to compensate for the loss of the network. However, an error-free PCIe5.0 data rate transmission can be obviously satisfied at an 8 × 8 NOCS, which means the net is compatible with major OCS applications. The top-speed eye-diagrams of the NOCS at 2 × 2, 4 × 4, and 8 × 8 port-scales are present in Figs. 4g, h, and i, respectively. The eye height was limited by the 3-dB bandwidth of the Mach-Zehnder modulator (MZM) of 45 GHz and that of the driver of 55 GHz. The side-eye of the 8 × 8 NOCS almost closed owing to the lower SNR caused by its high loss. Detailed experimental and simulation methods, as well as a discussion on the performance potential of the NOCS, can be found in Supplementary Section 8.

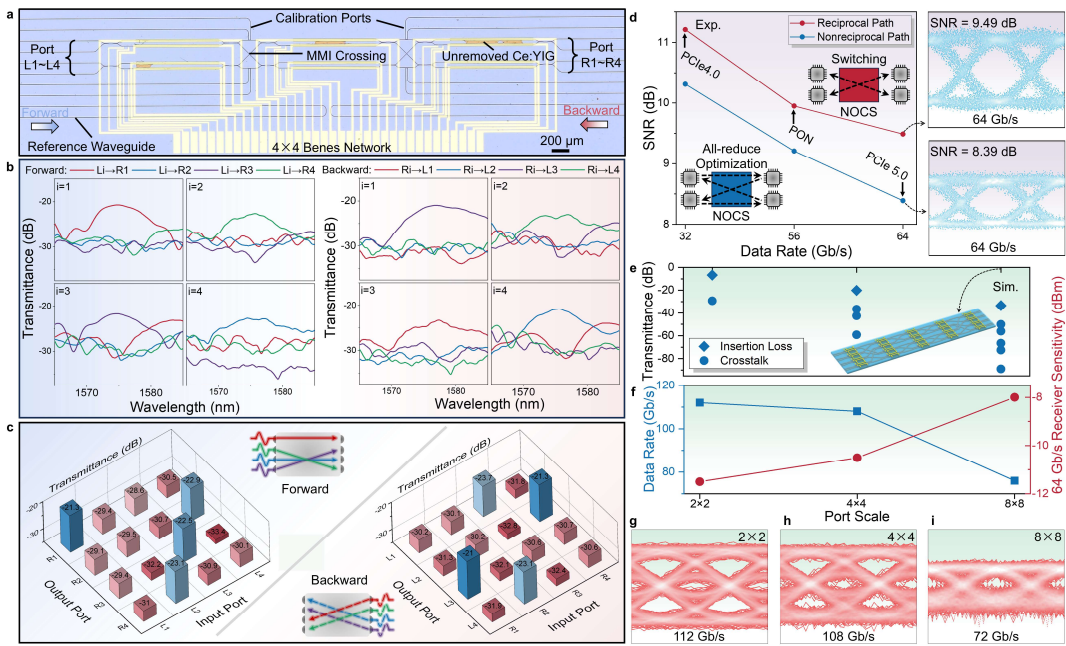

**Fig. 4 Experimental results of the MO NOCS networks. a**, Optical micro photo of the fabricated 4 × 4 MO nonreciprocal network. (**b**) Forward and backward transmission spectra and (**c**) transmission matrices of the nonreciprocal OCS. The inset shows the schematic of the routing configurations. **d**, Experimental measurement of SNR and eye-diagrams of a 2 × 2 NOCS under different data rates. Simulated (**e**) IL & crosstalk, (**f**) top-speed data rate and 64 Gb/s receiver sensitivity, and (**g**–**i**) eye-diagrams of the NOCS at 2 × 2, 4 × 4, and 8 × 8 port-scales.

## 5. Discussion

The network operation logic verified by the experiment corresponds to the application scenario in Fig. 1f. In AI data centers, several GPUs are often fully connected to achieve higher compute density, forming scale-up networks[47, 48]. However, the scalability of these networks is inherently limited by the fan-out capability of individual GPUs[49]. For example, to established an 8-GPU fully connection, each GPU requires at least seven fan-out ports. As a consequence, the bandwidth density of a GPU chips determines the maximum achievable scale of a fully interconnected network. OCS offers a promising solution to overcome this limitation by enabling more GPUs to join a scale-up network without significantly increasing latency or power consumption[5, 49]. NOCS, as demonstrated in this work, provides additional functionalities that surpass those of conventional OCS. A key advantage of NOCS is its ability to enable bidirectional, full-duplex operations. This is particularly beneficial in scenarios such as an all-reduce operation[5, 49], where a signal received from one GPU must be simultaneously sent to another. Conventional OCS schemes typically rely on time division, wavelength division, or space division multiplexing, which are often limited by latency, system complexity, or the number of fan-out ports. In contrast, NOCS can realize an all-reduce type operation without introducing additional overhead. Its nonreciprocal properties allow simultaneous uplink and downlink communication within the same connection path, eliminating the need for complex multiplexing strategies. This capability can significantly improve the efficiency of GPUs interconnect communication and reduce the hardware requirements for achieving full connectivity.

Another critical advantage of NOCS is its ability to dynamically reconfigure the connection topology according to the training tasks and models requirements during AI training in data center. During AI training, the communication topologic between GPUs often change depending on the specific model and optimization algorithm being used. NOCS enables ring-based all-reduce operations with any number of nodes in any direction simultaneously, which provides enhanced flexibility and efficiency in GPU computing clusters.

Not only can this OCS support scale-up networks, but it can also play a critical role in scale-out networks, such as in the spine layer or even the aggregation layer, enabling full-duplex communication between uplink and downlink to improve communication efficiency[3, 50]. Similarly, it can be extended to interconnections among multiple data centers or connections between data centers and base stations.

Beyond data centers, NOCS holds significant promise for ISAC in the upcoming 6G era. In this case, the sensing signal is received from the user end and processed in the local centralized unit/distributed unit (CU/DU). The communication signal is transmitted and received through the user end, active antenna unit (AAU), CU/DU, and data center or core net[50, 51], as shown in Fig. 1f. In practice, different CUs and DUs at different places are connected to jointly process the communication and sensing signals. The downlink communication signal is from core net or datacenter to the user end, and the sensing signal is usually received from user end to local or nearby CU/DU. In traditional OCS systems, the downlink communication signal and the uplink sensing signal are transmitted with time division since the analog link is symmetric. However, the communication signal occupies most of the time slot due to user demand. Thus, NOCS is a candidate solution to achieve ISAC signal transmission without affecting communication quality. By enabling simultaneous, independent transmission of downlink communication and uplink sensing signals within the same channel, the NOCS seamless integrated sensing and communication functionalities, which is critical for applications such as autonomous vehicles, smart cities, and next-generation wireless networks.

## 6. Conclusion

We have performed the first MO heterogeneous integrated OCS network that achieves nonreciprocal capabilities, addressing key limitations of traditional OCS architectures. By leveraging the silicon photonics platform and cascading MZI-based MO switches into a 4 × 4 Benes network, we achieved flexible reconfiguration and bidirectional operation. The independent control of forward and backward transmission states of each switch unit offering functionalities such as isolation, multi-port circulation, and direction-dependent routing. The fabricated MO switches exhibited nanosecond-scale MO reconfiguration speeds and expected phase shifting capability.

To overcome the challenges of integrating MO materials, we developed an innovative process for the precise patterning of Ce:YIG and YIG films, which not only minimizes optical loss but also ensures compatibility with silicon photonic platforms.

This work demonstrates the feasibility and scalability of MO-based NOCS networks, enabling energy-efficient, low-latency interconnects for AI workloads and ISAC systems in next-generation 5G/6G networks. With further optimization of MO material properties, fabrication processes, and device performance, the proposed NOCS technology can pave the way for advanced optical networks with unprecedented functionality and scalability.

**Methods**

**Patterned integrations of Ce:YIG/YIG.** As shown in Extended Data Fig. 1, the device fabrication started with a SOI wafer consisting of a 220-nm-thick top silicon layer, a 2-μm-thick buried oxide layer, and a 400-μm-thick silicon substrate. First, a 200-nm-thick hard mask $SiO_2$ layer was deposited on the wafer surface through plasma enhanced chemical vapor deposition (PECVD, Oxford Instruments PlasmaPro 80 PECVD). The wafer was then diced into chips. A positive resist (ZEP520A), spin-coated at 6000 rpm, was exposed using an Elionix ELS-F125G8 electron beam lithography (EBL) system with a beam current of 1 nA. After developing in o-xylene, the pattern was transferred to the hard mask layer via inductively coupled plasma (ICP, Leuven HAASRODE-E200A) etching with $CHF_3$. The chip was cleaned in ultrasonic baths with N-methyl-2-pyrrolidone, acetone, and isopropanol to strip the residual resist. A second ICP dry etching with $SF_6$ and $CF_4$ gases was performed to form the Si WG, followed by hydrofluoric acid (HF) wet etching to remove the $SiO_2$ mask.

Then, the LOR 5A and AZ5214 resists were sequentially spin-coated on top of the patterned Si layer. To realize the patterned integrations of YIG, a maskless aligner (Heidelberg MLA150) was used to define the MO windows for YIG deposition. After exposure, the bilayer resist was developed using AZ 300 MIF and cleaned afterwards with deionized water. Following, a 50 nm-thick amorphous YIG film was deposited at room temperature onto the defined windows via magnetron sputtering. The low deposition temperature ensured the stability of the photoresist during this process. Subsequently, the device was immersed in an acetone solution, and a blanket lift-off step was performed to completely remove the amorphous YIG and photoresist from the non-MO phase shift regions. The remaining YIG film was then subjected to rapid thermal annealing (RTA) under a pure oxygen environment, which purified the film and formed a crystallization template for subsequent Ce:YIG growth. The same procedure was employed to produce a patterned Ce:YIG film with a thickness of ~100 nm. The patterned areas of YIG and Ce:YIG differed slightly, as the YIG MO section area was slightly increased by ~10 μm to ensure complete Ce:YIG crystallization.

Next, a 2-μm-thick $SiO_2$ layer, which was thick enough to protect the optical WG from metal absorption, was deposited on the chip surface by PECVD as an upper cladding. A bilayer resist (LOR 5A and AZ5214) was spin-coated, and a third MLA was performed to pattern the evaporation windows of Ti heaters on top of the $SiO_2$ cladding. After a development process, a 200-nm-thick titanium layer was subsequently deposited using electron beam evaporation (EBE, Denton Vacuum Explorer-14), followed by lift-off in acetone. The same procedure was used to define the evaporation windows of the gold wires. EBE was used to deposit a 20-nm-thick Ti adhesion layer and an 800-nm-thick gold layer, after which lift-off was carried out to form the gold electrodes and pads. Finally, the chip was cleaved to expose the facets for WG edge coupling.

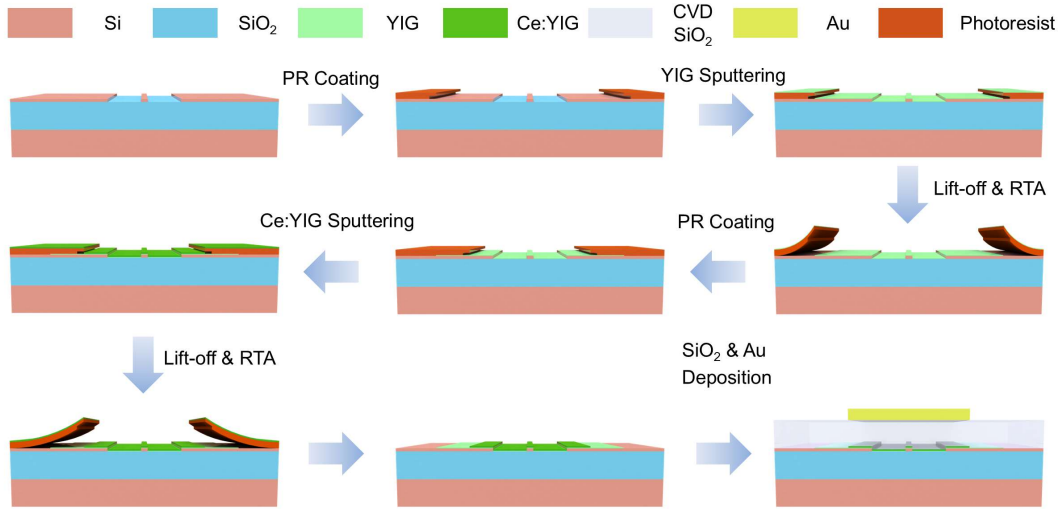

**Extended Data Fig. 1** Process flow of nonreciprocal phase shifters.

## Data availability

All data are available from the corresponding authors upon reasonable request.

## Acknowledgements

This work is supported by Zhangjiang Laboratory.

## Author contributions

S.L., L.B. and C.W. conceived the project. S.L., Z.T., and R.S. fabricated the samples and performed the measurements. Y.Y., T.Z. and D.W. developed the integration process of Ce:YIG/YIG films. Y.W. and F.H. performed the communication experiments and simulations. S.L. and P.Z. performed the data processing. C.Y. designed all the passive components. B.H. and H.C. provide the guidance and supervised the project. All the authors contributed to the discussions.

# Competing interests

The authors declare no competing interests.